\documentclass[12pt]{article}
\usepackage[final]{graphicx}
\usepackage{amsmath}
\usepackage{amssymb}
\usepackage{xcolor}
\usepackage{tensor}
\usepackage{braket}
\usepackage{units}
\usepackage[hypertex,colorlinks=true,linkcolor=red,citecolor=blue]{hyperref}

\usepackage{amssymb}
\newcounter{comment}

{\refstepcounter{comment}%
\begin{quote}
\ttfamily\small$\blacksquare$ \textbf{\underline{Comment} $\sharp$\thecomment:}}%
{\end{quote}}

{%\refstepcounter{comment}
\begin{quote}
\ttfamily\small$\blacktriangleright$ \textbf{\underline{Reply} $\sharp$\thecomment:}}%
{\end{quote}}

\newcommand{\tffF}{{\mathcal F}}
\newcommand{\tfftH}{\widetilde{\mathcal H}}
\newcommand{\tfftE}{\widetilde{\mathcal E}}

\newcommand{\pS}{{\mathrm{pS}}}

\newcommand{\req}[1]{(\ref{#1})}

\newcommand{\cQ}{{\cal Q}}

\newcommand{\xB}{x_{\rm B}}
\newcommand{\muR}{\mu_{\rm R}}
\newcommand{\muF}{\mu_{\rm F}}

\newcommand{\muphi}{\mu_{\varphi}}

\newcommand{\qmi}{{q^{(-)}}}
\newcommand{\umi}{u^{(-)}}
\newcommand{\dmi}{d^{(-)}}
\newcommand{\smi}{s^{(-)}}

\newcommand{\g}{{\text{G}}}

\newcommand{\GeV}{{\rm GeV}}

\def\muFgpd{\relax\ifmmode\mu_\text{F,GPD}^2\else{$\mu_\text{F,GPD}^2${ }}\fi}
\def\muFda{\relax\ifmmode\mu_\text{F,DA}^2\else{$\mu_\text{F,DA}^2${ }}\fi}
\def\muO{\relax\ifmmode{\mu_{0}^{2}}\else{$\mu_{0}^{2}${ }}\fi}
\def\Mev{\relax\ifmmode{\text{MeV}}\else{MeV{ }}\fi}

\def\bu{\overline{u} \,}
\def\bv{\overline{v} \,}

\def\CF{C_\text{F} }

\def\CA{C_\text{A} }

\def\Li{\relax\ifmmode{\text{Li}_{2}}\else{Li$_2${ }}\fi}
\def\im{\Im{\rm m}}

\font\cmss=cmss12 
\def\1{\hbox{{1}\kern-.25em\hbox{l}}}
\def\bfZ{\relax{\hbox{\cmss Z\kern-.4em Z}}}
\begin{document}

\begin{titlepage}

\centerline{\large \bf Next-to-leading order corrections}
%\centerline{\large \bf to deeply virtual meson leptoproduction processes}
\centerline{\large \bf to deeply virtual production of pseudoscalar mesons}

\vspace{10mm}

\centerline{G.~Duplan\v{c}i\'{c},
            D.~M\"{u}ller, and
            K.~Passek-Kumeri\v{c}ki
}

\vspace{8mm}

\vspace{4mm}
\centerline{\it Theoretical Physics Division, Rudjer Bo{\v s}kovi{\'c} Institute}
\centerline{\it HR-10002 Zagreb, Croatia}

\vspace{10mm}

\centerline{\bf Abstract}
{\ }

\noindent
We complete the perturbative next-to-leading order corrections
to the hard scattering amplitudes of deeply virtual meson leptoproduction processes at leading twist-two level
by presenting the results for the production of flavor singlet
pseudoscalar mesons.
The new results are given in the common momentum fraction representation
and in terms of conformal moments. We also comment on the flavor singlet results for deeply virtual vector meson production.

\vspace{0.5cm}

\noindent

\vspace*{12mm}


\vspace*{22mm}
\noindent

\end{titlepage}
\begin{figure}[t]
\begin{center}
\includegraphics[width=10cm]{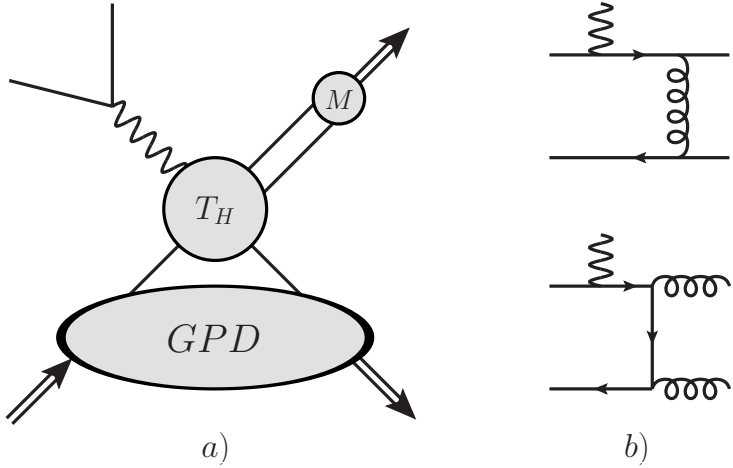}
\end{center}
\vspace{-5mm}
\caption{\small  a) Factorization of the DVMP amplitude for a longitudinally polarized photon exchange in GPD, meson distribution amplitude, and hard scattering part $T_H$. In b) representative LO diagrams for the hard scattering amplitude are shown for the quark--quark (up) and quark--gluon (down) channel.   }
\label{Fig-DVMP}
\end{figure}


\noindent
i.~Much experimental effort has been spent during the last decade and will be spent in future by the JLAB and COMPASS  collaborations to measure exclusive leptoproduction processes in the deeply virtual regime in which the virtuality of the exchanged photon is considered as large.  The phenomenological goal of such measurements is to access generalized parton distributions (GPDs) \cite{Mueller:1998fv,Radyushkin:1996nd,Ji:1996nm},
 which encode partonic information that are complementary to parton distribution functions or hadronic distribution amplitudes, see, e.g., Refs.\ \cite{Diehl:2003ny,Belitsky:2005qn}.  These process independent (universal) quantities are related to observables by convolution formulae where the hard--scattering amplitude is perturbatively calculable in leading twist--two approximation. Examples of such observables are the transverse cross section of deeply virtual Compton scattering (DVCS) and the longitudinal cross sections for the deeply virtual meson production (DVMP)  of pseudo scalar and longitudinally polarized vector mesons. They are  experimentally accessible in
exclusive lepton--nucleon reaction $l(k) N(P_1) \to l(k^\prime) N(P_2)  M(q_2)$
in which the virtual one-photon exchange contribution with four momentum $q_1=k-k^\prime=P_2 +q_2 -  P_1$ is the dominant one.
To utilize the factorization theorem \cite{Collins:1996fb}, it is required to address the longitudinally polarized differential cross section
\cite{Mankiewicz:1997aa,Mankiewicz:1997uy,Mankiewicz:1998kg}, e.g., in the notation of Ref.\ \cite{Mueller:2013caa}
it is given as transition form factors (TFFs) that appear in a form factor decomposition of the  amplitude. For example, in the case of pseudo scalar meson
production
\begin{eqnarray}
\label{tffF-def}
\epsilon_1^\mu(0) \langle M N| j_\mu |N\rangle \!\!\!&=&\!\!\!
%\left\{{
%\overline{u}(P_2,s_2) \bigg[
 %{\not\!\! m}\, \tffH_{M} + i \sigma_{\alpha \beta} \frac{m^\alpha \Delta^\beta}{2 M_N}\, \tffE_{M} \bigg]u(p_1,s_1)
%\quad \mbox{parity even}
 %\atop
 e\, \overline{u}(P_2,s_2) \bigg[
 {\not\!\! m} \gamma_5 \,  \tfftH_{M} + \gamma_5 \frac{m\cdot(P_2-P_1)}{2 M_N}\, \tfftE_{M} \bigg]u(P_1,s_1)\,,
  %\quad \mbox{parity odd}
 %}\right. ,
\end{eqnarray}
where the vector $m^\mu$ might be equated to $(q_1+q_2)^\mu/(P_1+P_2)\cdot(q_1+q_2)$ and $e$ is the unit electrical charge. The TFFs, generally denoted as $\tffF_{M}(\xB,t,\cQ^2)$,  depend on the Bjorken variable $\xB = \cQ^2/2 P_1\cdot q_1$, the momentum transfer square $t=(P_2-P_1)^2$, and the photon virtuality square $ \cQ^2 = - q_1^2$.
 The leading order formalism for different channels of such processes, depicted in Fig.~\ref{Fig-DVMP}, were worked out for some time \cite{Frankfurt:1995jw,Radyushkin:1996ru,Frankfurt:1997fj,Mankiewicz:1997uy,Mankiewicz:1997aa,Mankiewicz:1998kg,Frankfurt:1999xe,Frankfurt:1999fp,Vanderhaeghen:1999xj}.

For setting up a robust GPD phenomenology there is necessity to address
perturbative higher--order as well as higher--twist corrections.
 The former ones can be calculated according to the state of the art
while the evaluation of higher twist corrections is a problematic task,
pioneered for DVCS by V.~Braun and A.~Manashov  \cite{Braun:2011dg,Braun:2011zr}.
Note that a fixed order calculation induces a residual scale dependence
that is maximal in the leading--order (LO) approximation.
To reduce this dependence it is necessary to take higher order corrections
into account.
DVMP for flavor non-singlet pseudo--scalar mesons and longitudinally polarized
vector mesons were already worked out at next-to-leading order (NLO)
level in Refs.\ \cite{Belitsky:2001nq} and \cite{Ivanov:2004zv}, respectively.
The NLO corrections of the former ones might be obtained by analytic continuation
from the existing result of the pion form factor,
see e.g., Ref.~\cite{Melic:1998qr}, while the latter one requires
a diagrammatic calculation of hard partonic processes.

In this study we address the NLO corrections for DVMP of
flavor singlet pseudoscalar mesons.
We calculate NLO corrections to the corresponding partonic processes
in the quark-quark channel $\gamma_L^{\ast} q \to (q \bar{q}) q$
and in the quark-gluon channel $\gamma_L^{\ast} q \to(g g) q$,
which was found to vanish at LO \cite{Kroll:2002nt,Baier:1982vlv}.
That  completes the compendium of NLO results for DVMP
at twist-two level.  We present our new results also in terms of conformal moments,
which allow to set up efficient GPD models and numerical code
for the analysis of experimental data.
In presenting our results we follow closely the notation
%of our previous work on vector meson production \cite{Mueller:2013caa},
of our previous work \cite{Mueller:2013caa}
and refer there for common definitions.
\\

%%%%%%%%%%%%%%%%%%%%%%%%%%%%
%%%%%%%%%%%%%%%%%%%%%%%%%%%%%

\noindent
ii.~According to the flavor content of the meson, the TFFs (\ref{tffF-def}) might be decomposed in partonic TFFs.
In particular, for the flavor octet  and singlet components  of the $\eta$ meson,
\begin{eqnarray}
|\eta^{(8)}\rangle
%&\!\!\!=\!\!\!&
= \frac{1}{\sqrt{6}}
\left(|u\bu\rangle + |d\bar{d}\rangle
-2 |s\bar{s}\rangle\right),
\quad
|\eta^{(0)}\rangle
%&\!\!\!=\!\!\!&
=\frac{1}{\sqrt{3}}
\left(|u\bu\rangle + |d\bar{d}\rangle
+ |s\bar{s}\rangle\right),
\end{eqnarray}
 we utilize the decompositions
\begin{eqnarray}
\tffF_{\eta^{(8)}}
% &\!\!\!=\!\!\!&
=
\frac{2}{3 \sqrt{6}} \tffF^{\umi}_{\eta^{(8)}}
- \frac{1}{3 \sqrt{6}}  \tffF^{\dmi}_{\eta^{(8)}}
+ \frac{2}{3 \sqrt{6}}  \tffF^{\smi}_{\eta^{(8)}},
%\label{tffF_eta8}\\
\quad
\tffF_{\eta^{(0)}}
% &\!\!\!=\!\!\!&
=
\frac{2}{3 \sqrt{3}} \tffF^{\umi}_{\eta^{(0)}}
- \frac{1}{3 \sqrt{3}}  \tffF^{\dmi}_{\eta^{(0)}}
- \frac{1}{3 \sqrt{3}}  \tffF^{\smi}_{\eta^{(0)}}
%\quad\mbox{with}\quad \tffF\in\{\tfftH, \tfftE\},
\label{tffF_eta0}
\end{eqnarray}
where $ \tffF\in\{\tfftH, \tfftE\}$ introduced in \req{tffF-def},
and the charge factors are included in
\req{tffF_eta0}.
These TFFs allow to address the corresponding charge odd quark GPDs
\begin{eqnarray}
\label{Deltaqmi}
F^\qmi(x, \eta, t,\mu^2) = F^q(x, \eta, t,\mu^2) - F^q(-x, \eta, t,\mu^2)\quad\mbox{for}\quad F \in \{\widetilde H, \widetilde E\},
\end{eqnarray}
which depend on the momentum fraction $x$, the skewness  $\eta$, $t$, and the renormalization scale $\mu$.
They are antisymmetric in $x$ and are thus assigned with a signature factor
$\sigma=+1$
($F^{q,(\sigma)}(-x,\eta,t)=-\sigma F^{q,(\sigma)}(x,\eta,t)$).
Our definitions, see, e.g., appendix A1 of Ref.~\cite{Mueller:2013caa}, are such that in the forward limit  $\widetilde H^\qmi$  reduces to the difference of standard  polarized quark ($\Delta q$) and anti-quark  ($\Delta \overline{q}$) distributions:
$
\widetilde H^{\qmi}(x,\eta=0,t=0,\mu^2) = \Delta q(x,\mu^2)-\Delta \overline{q}(x,\mu^2)
\quad\mbox{for}\quad x>0.
$
The  $\widetilde H^{\qmi}$ and $\widetilde E^{\qmi}$ GPDs   satisfy the
evolution equation
\begin{eqnarray}
\mu^2 \frac{d}{d\mu^2} F^{\qmi}(x,\xi,t,\mu^2) = \int_{-1}^1\!\frac{dy}{2\xi}\, {^+V}\left(\frac{x+\xi}{2\xi},\frac{y+\xi}{2\xi},\alpha_s(\mu)\right) F^{\qmi}(y,\xi,t,\mu^2)
\,\,\mbox{for}\,\, F \in \{\widetilde H, \widetilde E\}\,.
\end{eqnarray}
The  kernel ${^+V} = \frac{\alpha_s}{2\pi} V^{(0)} + \frac{\alpha_s^2}{(2\pi)^2} {^+V}^{(1)} + O(\alpha_s^3)$ is in LO approximation given by
\begin{eqnarray}
\label{V^{(0)}}
 V^{(0)}(u, v)= \CF\, \theta\!\left(\!1 - \frac{u}{v}\!\right)\theta\!\left(\!\frac{u}{v}\!\right) {\rm sign}(v) \frac{u}{v}\left[ 1 + \frac{1}{(v-u)_+}\right]
+ \frac{3\CF}{2} \delta(u-v)
+ \left\{ {u \to \bu \atop  v \to \bv }  \right\}
\, ,
\end{eqnarray}
where $\CF=4/3$, $\bu=1-u$, and $\bv=1-v$.
The NLO kernel can be found in Eq.~(177) of Ref.\ \cite{Belitsky:1999hf}, denoted there as ${^{QQ} V}^{(1)+}$.

The formation of the meson is described by a  distribution amplitude (DA), see Fig.~\ref{Fig-DVMP}.
In the DV$\eta^{(0)}$P process it belongs to the flavor singlet sector and might be presented by a   vector
 \begin{eqnarray}
 \label{DA-singlet}
\mbox{\boldmath $\varphi$}_{\eta^{(0)}}(v,\mu^2) = \left({\varphi^\Sigma_{\eta^{(0)}}(v,\mu^2) \atop \varphi^\g_{\eta^{(0)}}(v,\mu^2)}\right)\,,\qquad  \varphi^\Sigma_{\eta^{(0)}}(\bv) = \varphi^\Sigma_{\eta^{(0)}}(v)\,, \quad \varphi^\g_{\eta^{(0)}}(\bv)= -\varphi^\g_{\eta^{(0)}}(v)
\end{eqnarray}
that contains the quark and gluon component, depending on the momentum fraction $v$ and the factorization scale $\mu$. The quark component is normalized as $\int_0^1 dv \varphi^\Sigma_{\eta^{(0)}}(v,\mu^2) =1$. More precisely, the entries of the flavor singlet meson DA (\ref{DA-singlet}) are defined by the following expectation values
\begin{eqnarray}
i f_{\eta^{(0)}}\varphi^\Sigma_{\eta^{(0)}}(v,\mu^2)  &\!\!\!= \!\!\!&\int \frac{d\kappa}{\pi} e^{i (v-\bv) (p\cdot n)  \kappa}
\sum_{q=u,d,s} \langle 0 | \overline{q}(-\kappa n)  n\cdot\gamma \gamma^5 q(\kappa n) |  \eta^{(0)}(p)\rangle_{(\mu^2)}
\\
i f_{\eta^{(0)}}\varphi^\g_{\eta^{(0)}}(v,\mu^2)  &\!\!\!= \!\!\!& \frac{2}{p\cdot n} \int \frac{d\kappa}{\pi} e^{i (v-\bv)  (p\cdot n)   \kappa}
\langle 0 | G^{+\mu}(-\kappa n)  i \epsilon_{\mu\nu}^\perp  G^{\nu+}(\kappa n) |  \eta^{(0)}(p)\rangle_{(\mu^2)}\,,
\end{eqnarray}
where $f_{\eta^{(0)}}$ is the decay constant.
Here
$\epsilon^{\perp}_{\mu\nu} = \epsilon_{\mu\nu\alpha\beta}\, n^{\ast\alpha} n^\beta$
with $\epsilon^{0123}=1$ and
$n^\mu$ and $n^{\ast\mu}$ being light-like vectors
satisfying $n\cdot n^{\ast} =1$ and $a^+ \equiv a\cdot n$.
The evolution of the DA is governed by the equation
\begin{eqnarray}
\mu^2 \frac{d}{d\mu^2}\mbox{\boldmath $\varphi$}_{\eta^{(0)}}(u,\mu^2) = \mbox{\boldmath $V$}(u,v|\alpha_s(\mu))\stackrel{v}{\otimes}\mbox{\boldmath $\varphi$}_{\eta^{(0)}}(v,\mu^2)\,,
\end{eqnarray}
where the matrix valued LO expression of the flavor singlet kernel is  \cite{Kroll:2002nt}
\begin{subequations}
\label{Vb^{(0)}}
\begin{eqnarray}
\label{bV^{(0)}-matrix}
\mbox{\boldmath $V$}(u, v|\alpha_s)
&=& \frac{\alpha_s}{2\pi}
\left(\!\!\!
\begin{array}{cc}
{^{\Sigma\Sigma}}V^{(0)}   &   {^{\Sigma\g}}V^{(0)}/2 \\
{2 ^{\g\Sigma}}V^{(0)}        &   {^{\g\g} V}^{(0)}
\end{array}
\!\!\!\right)\!\!
(u, v) + O(\alpha_s^2)\,,
\\
{^\text{AB} V}^{(0)}(u, v)
&=&
\theta(v - u )
\; {^\text{AB}}v^{(0)} (u,v) \pm
\left\{ {u \to \bar u \atop  v \to \bar v }\right\}
\mbox{\ for\ }
\left\{ {\text{A}=\text{B} \atop \text{A} \not= \text{B}}\right. .
\nonumber
\end{eqnarray}
The quark-quark entry ${^{\Sigma\Sigma}}V^{(0)}$  is given by the non-singlet kernel (\ref{V^{(0)}}) and the remaining entries are
\begin{eqnarray}
%\label{V-QQ}
%{^{\Sigma\Sigma}}v^{(0)} (u,v) &\!\!\! =\!\!\! & \CF\, \frac{u}{v}\left[ 1 + \frac{1}{(v-u)_+} + \frac{3}{2} \delta(u-v)\right]\,,
%\\
\label{V-QG,V-GQ}
{^{\Sigma\g}}v^{(0)} (u,v) &\!\!\! =\!\!\! &  - n_f\, \frac{u}{v^2}\,,
\qquad
{^{\g\Sigma}}v^{(0)}(u,v)  = \CF\, \frac{u^2}{v} \,,
\\
\label{V-GG}
{^{\g\g}}v^{(0)}(u,v)
&\!\!\! =\!\!\! & \CA\, \frac{u^2}{v^2}\left[2+ \frac{1}{(v-u)_+} \right] - \frac{\beta_0}{2} \delta(u-v)\,, \quad
\end{eqnarray}
\end{subequations}
where $\beta_0 =2/3 n_f - 11 \CA/3$ and $\CA=3$, and $n_f$ is the number of active quarks.
The NLO corrections to the evolution kernels are presented in Eqs.~(177)--(181) of Ref.~\cite{Belitsky:1998gc}.

The partonic TFFs (\ref{tffF_eta0}) are predicted to leading twist-two accuracy by the convolution formula
\begin{eqnarray}
\label{tffFeta0}
\tffF^{\qmi}_{\eta^{(0)}}(\xB,t,\cQ^2)
&\!\!\! \stackrel{\rm tw-2}{=}\!\!\! &
\frac{4\pi C_F f_{\eta^{(0)}}}{N_c \cQ}
 \int_{-1}^1\! \frac{dx}{2\xi}  \int_{0}^1\! dv
 F^{\qmi}(x,\xi,t,\muF^2)
 \\
 &&\phantom{\frac{C_F f_{\eta}}{N_c \cQ}
 \int_{-1}^1\! \frac{dx}{2\xi}  \int_{0}^1\! dv}\times
\mbox{\boldmath $T$}\!\left(\!\frac{\xi+x-i \epsilon}{2(\xi-i \epsilon)},v\Big|\alpha_s(\muR),
\frac{\cQ^2}{\muF^2},\frac{\cQ^2}{\muphi^2},\frac{\cQ^2}{\muR^2}\right)
\mbox{\boldmath $\varphi$}_{\eta^{(0)}}(v,\muphi^2),
\nonumber
\end{eqnarray}
where $\xi \simeq \xB/(2-\xB)$, the number of colors is $N_c =3$,   and
 $\mbox{\boldmath $T$}(u,v|\cdots) = \alpha_s \left( \frac{1}{\bu \bv}, 0\right) + O(\alpha_s^2),$  i.e., the gluonic component vanishes in LO approximation. Note  that the factor $4\pi$  in the overall normalization was reshuffled in \cite{Mueller:2013caa}.

Let us add that the results for DV$\eta^{(8)}$P TFFs formally follows from (\ref{tffFeta0})  by reduction to the flavor non-singlet case, i.e., we set
$\mbox{\boldmath $\varphi$}_{\eta^{(0)}}(v,\muphi^2) \to \varphi_{\eta^{(8)}}(v,\muphi^2)$ and $\mbox{\boldmath $T$} \to {^{+}T} $,
where
\begin{eqnarray}
\label{T-expand}
{^{+}T}(u,v|\cdots) &\!\!\!=\!\!\!& \alpha_s(\muR) T^{(0)}(u,v)  + \frac{\alpha^2_s(\muR)}{2\pi}  {^{+}T}^{(1)}\!\!\left(\! u,v\Big|\frac{\cQ^2}{\muF^2},\frac{\cQ^2}{\muphi^2},\frac{\cQ^2}{\muR^2}\!\right) + O(\alpha_s^3)\,,
 \end{eqnarray}
with $T^{(0)}(u,v)= 1/\bu\bv$.
The NLO expression for ${^{+}T}^{(1)}$ is presented in
Eqs.~(4.39) and (4.41) of  Ref.\ \cite{Mueller:2013caa}, where the signature factor is $\sigma =+1$.
\\

\noindent
iii.~ The hard scattering amplitude of the partonic processes $\gamma_L^{\ast} q(p_1)  \to [q(v) \bar{q}(\bv)] q(p_2)$  and $\gamma_L^{\ast} q(p_1)  \to [g(v) g(\bv)] q(p_2)$ are calculated in the collinear approximation, where  the incoming [outgoing] quark  GPD momentum is $p_1 = (x+\xi) P/2$  [$p_2 = (x-\xi) P/2$] with $P=P_1+P_2$ and the quark [anti-quark] momentum of the meson is $v q_2$  [$\bv q_2$]. In the calculation we employed dimensional regularization together with the $\gamma^5$-prescription of  t` Hooft--Veltman, equivalent to Breitenlohner-Maison prescription \cite{HooVel72,BreMai77}. In this HVBM scheme one renders a mathematically consistent result.
 Based on the one-loop Feynman integral reduction formalism \cite{Duplancic:2003tv}, the  regularized hard scattering amplitude in $D$ dimensional space
\begin{eqnarray}
\overline{\mbox{\boldmath $T$}}(u,v|\alpha_s,\cdots) = \alpha_s \overline{\mbox{\boldmath $T$}}^{(0)}(u,v) + \frac{ \alpha_s^2}{2\pi} \overline{\mbox{\boldmath $T$}}^{(1)}(u,v|\cdots)
\quad\mbox{with}\quad
\overline{\mbox{\boldmath $T$}}^{(0)}=
\left (\frac{6-D}{2}\frac{1}{\bu\bv},0\right)
\end{eqnarray} was calculated and cross checked at one loop level by two independently written codes. In one Feynman diagrams were implemented by hand and in the other generated with the FeynArt program \cite{Hahn:2000kx}. The collinear singularities were regularized by taking  $D=4+2\epsilon$
and they were absorbed in the dressed meson DA and GPD 
via the {\em modified} minimal subtraction scheme.
Note that due to vanishing LO, the NLO gluon and pure singlet quark contributions are ultraviolet finite.
The dressed hard scattering amplitude is finally obtained by taking the limit
$$
\mbox{\boldmath $T$}(u,v|\alpha_s,\cdots) = \lim_{D\to 4}\int_0^1\!du^\prime\!\!\int_0^1\!dv^\prime \; Z(u^\prime,u|\alpha_s)
\overline{\mbox{\boldmath $T$}}(u^\prime,v^\prime|\alpha_s,\cdots) \mbox{\boldmath $Z$} (v^\prime,v|\alpha_s)  \,,
$$
where the $Z$-factors  to one loop order accuracy,
 expressed by the  kernels  (\ref{V^{(0)}}) and (\ref{Vb^{(0)}}), are
\begin{subequations}
\label{Zfactors}
\begin{eqnarray}
Z(u,v)  &\!\!\!=\!\!\!&  \delta(u-v) + \frac{2 \left(4 \pi  e^{-\gamma_E}\right)^{\frac{4-D}{2}}}{4-D} \frac{\alpha_s}{2\pi} V(u, v) + O(\alpha_s^2)\,,
\\
\mbox{\boldmath $Z$}(u,v)  &\!\!\!=\!\!\!&
\left(
  \begin{array}{cc}
    \delta(u-v) & 0 \\
    0 &   \delta(u-v) \\
  \end{array}
\right) + \frac{2 \left(4 \pi  e^{-\gamma_E}\right)^{\frac{4-D}{2}}}{4-D} \frac{\alpha_s}{2\pi}  \mbox{\boldmath $V$}^{(0)}(u, v) + O(\alpha_s^2)\,,
\end{eqnarray}
\end{subequations}
with renormalized $\alpha_s$ and
$\gamma_E = 0.5772\dots $ is the Euler-Mascheroni constant.

To transform from the HVBM scheme to the common adopted one, requiring that the spin independent and spin dependent  evolution kernels in the flavor non-singlet case are the same, in addition to the minimal subtraction  a finite subtraction should be performed with the $z$-factor
\begin{eqnarray}
\mbox{\boldmath $z$}^{\rm HVBM}(u,v)&\!\!\!=\!\!\!&\left(
  \begin{array}{cc}
    \delta(u-v) & 0 \\
    0 &   \delta(u-v) \\
  \end{array}
\right)
+ \frac{\alpha_s}{2\pi} \left(
  \begin{array}{cc}
   4 \CF V^a(u,v) & 0 \\
    0 & 0 \\
  \end{array}
\right)  + O(\alpha_s^2)\,,
\end{eqnarray}
where $V^a(u,v) = \theta(v-u)\frac{u}{v} + \theta(u-v)\frac{\bu}{\bv}$.
This scheme transformation does not affect the quark-gluon channel and
contributes to the flavor non-singlet part, which is already known
\cite{Melic:2001wb}.
Note that this is entirely in agreement with the definition used
in deep inelastic scattering, see, e.g.,
Eqs.\   (33) --(39)  and (40) in Ref.\ \cite{Vogelsang:1996im}, where the correspondence $4 \CF V^a(u,v)  \leftrightarrow 4\CF (1-z)$ holds.

The NLO corrections to the hard scattering amplitude of DV$\eta^{(0)}$P,
\begin{subequations}
\label{T5-expand}
\begin{eqnarray}
\mbox{\boldmath $T$}(u,v|\cdots)= \left(^\Sigma T(u,v|\cdots),\frac{n_f}{\CF}\, {^\g T}(u,v|\cdots)\right),
{^\Sigma T}(\cdots) = T(\cdots) +n_f {^\pS T}(\cdots),
\end{eqnarray}
contains besides $T$, see Eq.~(\ref{T-expand}),
 the pure singlet ($\pS$) quark and the gluonic ($\g$) entries,
\begin{eqnarray}
%T(u,v|\cdots) &\!\!\!=\!\!\!& \alpha_s(\muR) T^{(0)}(u,v)  + \frac{\alpha^2_s(\muR)}{2\pi}  T^{(1)}\!\!\left(\! %u,v\Big|\frac{\cQ^2}{\muF^2},\frac{\cQ^2}{\muphi^2},\frac{\cQ^2}{\muR^2}\!\right) + O(\alpha_s^3)\,,
%\\
% T^{(0)}(u,v)&\!\!\!=\!\!\!&  \frac{1}{\bu\bv} \,,
%\\
{^\pS T}(u,v|\cdots)&\!\!\!=\!\!\!& \frac{\alpha^2_s(\muR)}{2\pi}\; {^\pS T}^{(1)}(u,v) +   O(\alpha_s^3)\,,
\\
{^\g T}(u,v|\cdots)  &\!\!\!=\!\!\!&  \frac{\alpha^2_s(\muR)}{2\pi} \left[
\CF {^\g T}^{(1,F)}\!\!\left(\!u,v\Big|\frac{\cQ^2}{\muphi^2}\!\right) + \CA {^\g T}^{(1,A)}(u,v)\right] + O(\alpha_s^3)\,.
\end{eqnarray}
\end{subequations}
Here, we exploit symmetry so that our NLO  expressions have only poles at $u=1$ and $[1,\infty]$ cuts on the positive real axis  in the complex $u$-plane:
\begin{subequations}
\label{pSTgT}
\begin{eqnarray}
{^\pS T}^{(1)}&\!\!\!=\!\!\!&
 \frac{\Li(v)-\zeta_2}{\bu \bv}-\frac{\ln\bv+\Li(v)}{\bu v}-\left[\frac{\vec{\partial}}{\partial v}\;v -2\right]\left[\frac{L(u,v)}{u(u-v)}\right]^{\rm sub} -\left[\frac{L(u,v)}{u(u-v)\bv}\right]^{\rm sub}
\\
{^\g T}^{(1),F}&\!\!\!=\!\!\!&
\frac{\ln\bv}{2\bu v^2} \left[\ln\frac{\cQ^2}{\mu_\varphi^2}-\frac{3}{2} +\frac{1}{2} \ln \bv \right]
-\frac{\ln\bu-u }{2u\bu}\, \frac{\ln\bv}{v\bv} -\frac{\Li(u)}{2u\bv}-\frac{\Li(u)-\zeta_2}{2\bu\bv}
\\
&&+
\frac{1}{4}\left[ \frac{\vec{\partial}^2}{\partial v^2}\, v\bv + 2 \right]\left[\frac{L(u,v)}{u(u-v)\bv}\right]^{\rm sub}
\,,
\nonumber\\
{^\g T}^{(1),A}&\!\!\!=\!\!\!&
\frac{\ln\bu}{4u\bu}\,\frac{\ln\bv}{v^2\bv}+\frac{\Li(u)}{2u\bu \bv}-\frac{\bv-v}{4\bu v\bv}\,\frac{\ln\bv}{ v}+
\frac{(\bv-v)\left[\Li(v)- \zeta_2\right]}{4\bu \bv^2}-\frac{(\bv-v)\Li(v)}{4\bu v^2}
\nonumber\\
&& -\frac{\bv-v}{4}\frac{\vec{\partial}}{\partial v}\left[\frac{L(u,v)}{u(u-v)\bv}\right]^{\rm sub},
\end{eqnarray}
\end{subequations}
where  $\zeta_2 = \pi^2/6$. The non-separable terms are expressed by end-point subtracted building blocks
\begin{subequations}
\label{L(u,v)}
\begin{eqnarray}
\left[
\frac{L(u,v)}{u(u-v)}
\right]^{\rm sub} &\!\!\! \equiv \!\!\!&
\frac{L(u,v)}{u(u-v)} +\frac{L(u=0,v)}{u v}\,,
\label{eq:Lsub-u1}
\\
\left[
\frac{L(u,v)}{u(u-v)\bv}
\right]^{\rm sub} &\!\!\! \equiv \!\!\!&
\frac{L(u,v)}{u(u-v)\bv} +\frac{L(u,v=1)}{u\bu\bv}+\frac{L(u=0,v)}{u v\bv}-\frac{L(u=0,v=1)}{u\bv}\,,
\end{eqnarray}
\end{subequations}
with $L(u,v) = \Li(u) -\Li(v)  +\ln \bu \ln v   -\ln\bv \ln v$.

The substraction of end-point singularities in the non-separable terms (\ref{L(u,v)}) ensures that they provide numerically small contributions. In the pure singlet quark result the most singular contribution is given by the pole $1/\bu$ at $u=1$.
Its residue is  a rather harmless function in $v$ that contain no end-point singularities. Thus, these perturbative corrections are relatively small. Contrarily, in the quark-gluon
channel the most singular term  $(\ln\bu)/\bu (\ln\bv)/\bv$  can potentially provide large corrections, which, however, are numerically suppressed in the large $N_c$ limit. Nevertheless, besides a $(\ln\bu)/\bu (\ln\bv)/2v^2 \sim  (\ln\bu)/2 \bu \bv$ term,  the net result has also $1/\bu$  pole contributions. The most singular terms can be collected into
$$
\frac{{^\g T}^{(1)}}{\CF} \sim \frac{\alpha_s^2}{2\pi}\left[\ln\bu + \ln\bv +2\zeta_2-\frac{1}{2}+\ln\frac{\cQ^2}{\muphi^2} + \frac{1}{N_c^2-1}
\left\{\ln\bu  \ln\bv + \ln\bu+1+\zeta_2 \right\} \right] \frac{1}{2\bu \bv}
$$
and might provide in dependence on the gluonic $\eta^{(0)}$     DA a moderate or sizeable correction.

We also calculated the flavor singlet hard scattering amplitude for longitudinal vector meson production,  e.g., for DV$\rho^{(0)}_L $P.
The results from Ref.~\cite{Ivanov:2004zv} are obtained making an
average over two transverse gluon polarization states.
However, it  is standard PDF convention to take an average over $D-2$
transverse polarizations available to gluons in $D$ dimensions.
Thus, the dimensional regularized LO hard scattering amplitude changes:
$$
\overline{\mbox{\boldmath $T$}}^{(0)}=
\left (\frac{D-2}{2}\frac{1}{n_f}\frac{1}{\bu\bv},\frac{D-2}{2}\frac{1}{\CF \xi} \frac{1}{\bu \bv}\right)
 \Rightarrow
\overline{\mbox{\boldmath $T$}}^{(0)}=
\left (\frac{D-2}{2}\frac{1}{n_f}\frac{1}{\bu\bv},\frac{1}{\CF \xi} \frac{1}{\bu \bv} \right)
$$
and by the same overall factor $2/(D-2)$ in the gluon entry at NLO (and beyond).   To ensure that the forward limit of the gluon GPD provides the common definition of the PDF, used in the phenomenology of (semi-)inclusive measurements, the original results \cite{Ivanov:2004zv} should be  corrected in the pure quark  singlet \cite{Ivanov:2004zv_erra}   and the gluon sector by an additional NLO term:
\begin{eqnarray}
\mbox{\boldmath $T$}^{(1)}(u,v|\cdots) \Rightarrow
\mbox{\boldmath $T$}^{(1)}(u,v|\cdots) +
\frac{1}{\bv}\int_0^1\!\!\frac{du^\prime}{\bu^\prime} \left(\frac{2}{\CF} ^{\g\Sigma}\!V^{(0)}(u^\prime,u), -\frac{1}{2n_f \xi}  ^{\Sigma\g}\!V^{(0)}(u^\prime,u)\right).
\end{eqnarray}
This change can be easily taken into account in the formula set of Ref.~\cite{Mueller:2013caa}  by the replacement
$$
\ln\frac{\cQ^2}{\muF^2} \Rightarrow \ln\frac{\cQ^2}{\muF^2}+1
\quad\mbox{and}\quad
\ln\frac{\cQ^2}{\muF^2} \Rightarrow \ln\frac{\cQ^2}{\muF^2}-1
$$
in ${^\pS T}^{(1)}$ [see Eqs.~(4.46a),  (4.47a), and (4.48a)  of Ref.~\cite{Mueller:2013caa} ] and in ${^\text{G}T}^{(1,F)}$ [see Eqs.~(4.51b), (4.52b), and (4.53b) of Ref.~\cite{Mueller:2013caa} ], respectively.
A more detailed account of here summarized NLO calculations,
as well as their application to other channels is in preparation \cite{DVMPlong}.
\\

\noindent
iv. For the GPDs we might employ a Mellin-Barnes integral representation  (for further details see Sec.~3.3 of Ref.~\cite{Mueller:2013caa}) and for the $\eta^{(0)}$ DA an integral conformal partial wave expansion.
 In such an expansion the evolution can be explicitly included in the TFFs (\ref{tffFeta0}), which read now as
 \begin{eqnarray}
\label{tffFeta0-MB}
\tffF^{\qmi}_{\eta^{(0)}}(\xB,t,\cQ^2)
&\!\!\! \stackrel{\rm tw-2}{=}\!\!\!&
\frac{4\pi C_F f_{\eta^{(0)}}}{N_c \cQ}
 \frac{1}{2 i} \int_{c- i \infty}^{c+ i \infty}\! dj\, \xi^{-j-1}
\left[
  i + \tan\!\left(\!\frac{\pi\,j}{2}\!\right)
  \right]
  \\
  &&\times
  \left[
   \sum^\infty_{k=0 \atop {\rm even}}
\mbox{\rm T}_{jk}\!\left(\cQ^2,\cQ_0^2\right)
\mbox{\boldmath $\varphi$}_{\eta^{(0)},k}(\cQ_0^2)
\right]
F_j^{\qmi}(\xi,t,\cQ_0^2)\,.
\nonumber
\end{eqnarray}
 The conformal GPD moments $F_j^{\qmi}(\xi,t,\cQ_0^2)$ at the input scale $\cQ_0$ coincide for integer $j=n$ with
\begin{eqnarray}
F_n^{\qmi}(\eta,t,\cQ_0^2)  = \frac{
\Gamma\big(\frac{3}{2}\big)\Gamma(n+1)}{2^{n}  \Gamma\big(n+\frac{3}{2}\big)}
\;\frac{1}{2}\int_{-1}^1\! dx\; \eta^{n} \, C_n^{3/2}\!\left(\!\frac{x}{\eta}\!\right)
F^{\qmi}(x,\eta,t,\cQ_0^2)\,,
\end{eqnarray}
 and those of the $\eta^{(0)}$-DA (\ref{DA-singlet}) are collected in the vector
 \begin{eqnarray}
 \mbox{\boldmath $\varphi$}_{\eta^{(0)},k}(\cQ_0^2)  &\!\!\!=\!\!\!&  \left({\varphi^\Sigma_{\eta^{(0)},k}(\cQ_0^2) \atop \varphi^\g_{\eta^{(0)},k}(\cQ_0^2)}\right)
 =
  \int_0^1\!dv \left({
   \frac{2(2 k+3)}{3(k+1)_2}C^{3/2}_k(v-\bv)\varphi^\Sigma_{\eta^{(0)}}(v,\cQ_0^2)
  \atop
  \frac{4(2 k+3)}{(k)_4}C^{5/2}_{k-1}(v-\bv)\varphi^\g_{\eta^{(0)}}(v,\cQ_0^2)
  }\right) ,
  \end{eqnarray}
 where $(k)_m = k\cdots (k+m-1)$ is the Pochhammer symbol and $C_k^\nu$ are the Gegenbauer polynomials of order $k$ and index $\nu$.
The zeroth moments are given by $\varphi^\Sigma_{\eta^{(0)},0}=1$ and $\varphi^\g_{\eta^{(0)},0}=0$ and, thus, the sum in the gluonic component always starts from $k=2$.

The  vector valued amplitude $\mbox{\rm T}_{jk}$ consist of the hard scattering one that is convoluted  with the evolution operators
\begin{eqnarray}
\label{eq:tffqF_M-MBI-b}
{\rm T}_{jk}(\cQ^2,\cQ_0^2) =  \sum^\infty_{l=0 \atop {\rm even}} \sum^\infty_{m=0 \atop {\rm even}}
 \mbox{\boldmath $T$}_{j+m,k+l}\!\left(\!\alpha_s(\muR),\frac{\cQ^2}{\muF^2},\frac{\cQ^2}{\muphi^2},\frac{\cQ^2}{\muR^2}\!\right)
 \mbox{\boldmath $E$}_{k+l,k}(\muphi,\cQ_0)\, {^+}E_{j+m,j}(\muF,\cQ_0)\,.
\end{eqnarray}
The evolution operator for the GPD moments, formally written as path ordered exponential
\begin{eqnarray}
{{^+}E}_{jm}(\mu,\mu_0) = {\cal P}\exp\left\{-\int_{\mu0}^\mu \frac{d\mu^\prime}{\mu^\prime}  {{^+}\gamma}_{jm}(\alpha_s(\mu^\prime)) \right\},
\end{eqnarray}
is expressed by the $\sigma=+1$ anomalous dimensions ${^+\gamma}_{jm}= \frac{\alpha_s}{2\pi} {\gamma}_{j}^{(0)} \delta_{jm} + \frac{\alpha_s^2}{(2\pi)^2}{^+\gamma}_{jm}^{(1)}  + O(\alpha_s^3)$ with
\begin{eqnarray}
\label{eq:gamma0}
\gamma_{j}^{{\rm }(0)}
= \CF \left( 4 S_{1}(j + 1) - 3 - \frac{2}{(j + 1 )( j + 2 )} \right),
\end{eqnarray}
where $S_1(n)=\sum_{m=1}^n \frac{1}{m}$ is the harmonic sum of order one.
The evolution operator,
\begin{equation}
\mbox{\boldmath $E$}_{km}(\mu,\mu_0)   = {\cal P}\exp\left\{-\int_{\mu0}^\mu \frac{d\mu^\prime}{\mu^\prime}  {\mbox{\boldmath $\gamma$}}_{km}(\alpha_s(\mu^\prime)) \right\},
\end{equation}
for the  $\eta^{(0)}$ DA is expressed by the anomalous dimension matrix  of conformal operators,
 \begin{equation}
 \label{gamma5}
  \mbox{\boldmath $\gamma$}_{km}= \frac{(2 k+3) (m+1)_2 }{ (2 m+3)(k+1)_2} \left(
\begin{array}{cc}
{\phantom{ \frac{6}{k (k+3)}} ^{\Sigma \Sigma}}\gamma_{km} & \frac{m (m+3)}{12}  {^{\Sigma {\rm G}}}\gamma_{km}\\
{ \frac{12}{k (k+3)} ^{{\rm G} \Sigma }}\gamma_{km} & \frac{m (m+3)}{k (k+3)} {^{{\rm G}{\rm G}}}\gamma_{km}
\end{array}\right),
 \end{equation}
where $ {^{AB}\gamma}_{km}= \frac{\alpha_s}{2\pi} {^{AB}\gamma}_{k}^{(0)} \delta_{km} + \frac{\alpha_s^2}{(2\pi)^2}{^{AB}\gamma}_{km}^{(1)}  + O(\alpha_s^3)$. To LO accuracy the quark-quark entry is given in (\ref{eq:gamma0}) and the three remaining entries read
\begin{subequations}
\begin{eqnarray}
\label{Def-LO-AnoDim-QG-V,Def-LO-AnoDim-GQ-V}
{^{\Sigma\g}\!\gamma}_{k}^{(0)} &\!\!\!=&\!\!\! -
\frac{12 n_f }{( k+ 1 )( k + 2 )}
\,,
\qquad
{^{\g \Sigma}\!\gamma}_{k}^{(0)} =
-\CF\frac{k (k+3)}{3( k + 1 )( k + 2 )}
\,,
\\
\label{Def-LO-AnoDim-GG-V} {^{\g\g}\!\gamma}_{k}^{(0)}
&\!\!\!=&\!\!\!  \CA \left(4S_1( k + 1 ) - \frac{8}{( k + 1 )( k + 2)}  \right)+ \beta_0
\,.
\end{eqnarray}
\end{subequations}
The  evolution operators are specified to NLO accuracy in Sec.~4.3 of Ref.~\cite{Kumericki:2007sa}, where, however, the anomalous dimension matrix (\ref{gamma5}) must be used.

 The conformal moments of the hard scattering amplitude (\ref{T5-expand}) read
 \begin{eqnarray}
 \mbox{\boldmath $T$}_{jk}(\cdots) = \frac{2^{j+1}\ \Gamma\big(j+\frac{5}{2}\big)}{\Gamma\big(\frac{3}{2}\big)\Gamma(j+3)}
  \left( 3\, {^\Sigma} c_{jk}(\cdots ) , \frac{3n_f}{\CF}\,{ ^\g c}_{jk}(\cdots ) \right),
  \quad
  {^\Sigma} c_{jk} =c_{jk}+n_f {^\pS } c_{jk}\,.
 \end{eqnarray}
 The integral values of the $c_{jk}$ coefficients are normalized as following
 \begin{subequations}
\begin{eqnarray}
{^A c}_{nk} &\!\!\!=\!\!\!& \int_0^1\!du\!\int_0^1\!dv\, 2 u\bu C_n^{3/2}(u-\bu)  {^A T}(u,v|\cdots)  2 v\bv C_k^{3/2}(v-\bv)\,,
\\
{^\g c}_{nk}&\!\!\!=\!\!\!&  \int_0^1\!du\!\int_0^1\!dv\,  2 u\bu C_n^{3/2}(u-\bu)  {^\g T}(u,v|\cdots)  12 v^2\bv^2 C_{k-1}^{5/2}(v-\bv)
\end{eqnarray}
\end{subequations}
for the quark-quark channel  $A\in\{q, \Sigma,\pS \}$ and the quark-gluon channel, respectively.

The perturbative expansion of these moments is analogous to those of the hard scattering amplitude (\ref{T5-expand}),  replace there ${^{\cdots} T}^{(1\dots)} (u,v|\cdots)$ by
${^{\cdots} c}^{(1\dots)}_{jk}(\cdots)$,
where  ${c}_{jk}^{(0)} =1$. The NLO expressions ${c}^{(1)}_{jk}$ for the quark-quark channel  can be read off from Eq.~(4.44) in Ref.\ \cite{Mueller:2013caa}, where  the signature is $\sigma=+1$. Utilizing the method and results presented in Sec.~4.1 of Ref.~\cite{Mueller:2013caa}, we find the remaining coefficients from the hard scattering amplitudes (\ref{pSTgT}):
\begin{eqnarray}
{^\pS c}_{jk}^{(1)}&\!\!\!=\!\!\!&
-\frac{(k+1)_2+2}{[(k+1)_2]^2}+\frac{\Delta\!S_2\!\big(\!\frac{k+1}{2}\!\big)+\Delta\!S_2\!\big(\!\frac{j+1}{2}\!\big)}{2}
+
\frac{ (k+1)_4}{2 k+3}\frac{\Delta\!S_2\!\big(\frac{1+j}{2},\!\frac{k+2}{2}\!\big)}{2} -\frac{ (k-1)_4}{2 k+3}\frac{\Delta\!S_2\!\big(\frac{1+j}{2},\!\frac{k}{2}\!\big)}{2}
\nonumber\\
\label{pSc_{jk}-odd}
\end{eqnarray}
for the pure singlet quark part and
\begin{eqnarray}
{^\g c}^{(1,F)}_{jk} &\!\!\!=\!\!\!&
-2S_1(j+1)\left[S_1(k+1)-1\right]+\frac{k(k+3)}{2(k+1)_2}\Bigg[\ln\frac{\cQ^2}{\mu^2_\varphi}-\frac{1}{2}-2 S_1(j+1)-2 S_1(k+1)
\nonumber\\
&&+\frac{1}{(k+1)_2}\Bigg]-
 \frac{ (k)_4}{2 k+3}\frac{(k+1) (k+4) \Delta\!S_2\!\big(\!\frac{j+1}{2},\frac{k+2}{2}\!\big)-(k-1) (k+2) \Delta\!S_2\!\big(\!\frac{1+j}{2},\frac{k}{2}\!\big)}{8}
\\
{^\g c}^{(1,A)}_{jk} &\!\!\!=\!\!\!&
S_1(j+1) \left[S_1(k+1)-1\right]+\frac{\zeta_2+1}{2}-\frac{2 (k+1)_2+2}{\left[(k+1)_2\right]^2}-\frac{ \Delta\!  S_2\! \big(\!\frac{j+1}{2}\!\big)}{4}-
\frac{(k+1)_2 -4}{2}
\nonumber\\
&&\times \frac{\Delta\! S_2\!\big(\!\frac{j+1}{2}\!\big)+\Delta\!S_2\!\big(\!\frac{k+1}{2}\!\big)}{4}
-\frac{(k)_4}{2 k+3}
\frac{(k+4) \Delta\! S_2\!\big(\!\frac{j+1}{2},\frac{k+2}{2}\!\big)+(k-1) \Delta\! S_2\!\big(\!\frac{j+1}{2},\frac{k}{2}\!\big)}{4}.
\end{eqnarray}
for the quark-gluon channel.
Here, $S_i(n)=\Sigma_{m=1}^n m^{-i}$ are the harmonic sums of order $i$ and
$$ \Delta\! S_2(n,m) = \frac{ \Delta\!S_2(n)- \Delta\!S_2(m)}{4 (n - m) (1 + 2 m + 2 n)}\,, \quad
\Delta\! S_2(n,n) =  -\frac{ \Delta\!S_3(n)}{2 (1+4 n)}
$$
with $\Delta\! S_i(n)=S_i(n)-S_i(n-1/2)$.

\begin{figure}[t]
\begin{center}
\includegraphics[width=16cm]{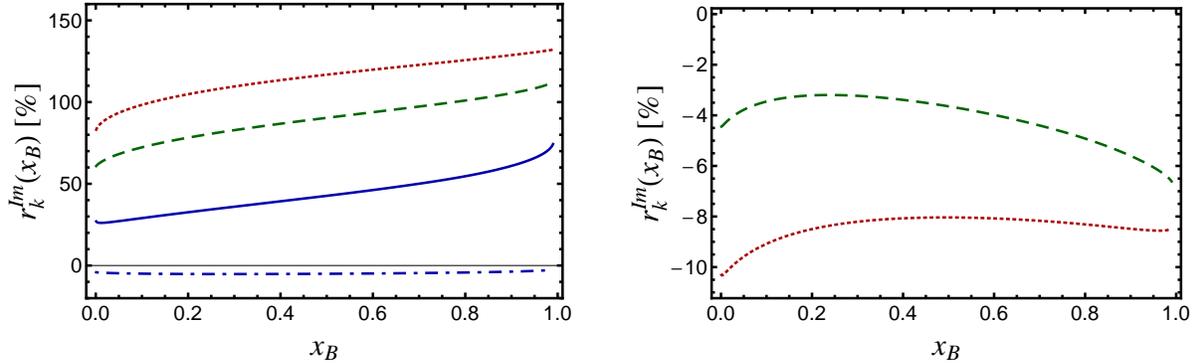}
\end{center}
\vspace{-5mm}
\caption{\small Relative NLO corrections (\ref{r_k^im}) to the imaginary part of the TFF (\ref{tffFeta0-MB}) versus $\xB$ for the  $k=0$ (solid), $k=2$ (dashed),  $k=4$ (dotted) partial waves arising from
the quark-quark channel (left panel) and quark-gluon channel (right panel). The pure singlet quark contribution for $k=0$ is shown as dash-dotted line in the left panel.  }
\label{Fig-NLO}
\end{figure}

To quantify the NLO corrections we take a simple model for the charge odd quark GPDs,
\begin{eqnarray}
F_j^{\qmi}(\xi,t=0,\cQ_0^2) = n^{\qmi}
\frac{6 \Gamma\left(j+\frac{1}{2}\right)}{\Gamma\left(j+\frac{9}{2}\right)}\left(\frac{\xi }{2}\right)^{j+1}\frac{ \Gamma\left(\frac{1}{2}\right)\Gamma)(j+2) }{\Gamma\left(j+\frac{3}{2}\right)}\; {_2F_1}\!\left({-j-1,j+2 \atop 1}\Big|\frac{-1+\xi }{2 \xi }\right),
\end{eqnarray}
which in the forward limit reduce to the PDF $F^{\qmi}(x,\xi=0,t=0,\cQ_0^2) =  n^{\qmi} x^{-1/2}(1-x)^3 $.  Setting $\alpha_s(\cQ_0 \sim 1.6 \GeV) =0.1 \pi$,  in Fig.\ \ref{Fig-NLO}  we show  the relative NLO corrections
\begin{eqnarray}
\label{r_k^im}
\mbox{\boldmath $r$}_k^{\im}(\xB,\cQ_0^2)=
\frac{ \frac{\alpha_s^2(\cQ_0)}{2\pi}
\im   \frac{1}{2 i} \int_{c- i \infty}^{c+ i \infty}\! dj\, \xi^{-j-1}
\left[
  i + \tan\!\left(\!\frac{\pi\,j}{2}\!\right)
  \right]
\mbox{\boldmath $T$}^{(1)}_{jk}\!\left(\cQ_0^2,\cQ_0^2\right)
F_j^{\qmi}(\xi,t=0,\cQ_0^2)
}{
\im \frac{1}{2 i} \int_{c- i \infty}^{c+ i \infty}\! dj\, \xi^{-j-1}
\left[
  i + \tan\!\left(\!\frac{\pi\,j}{2}\!\right)
  \right]
\mbox{\boldmath $T$}_{j{k=0}}\!\left(\cQ_0^2,\cQ_0^2\right) \left({1 \atop 0}\right)
F_j^{\qmi}(\xi,t=0,\cQ_0^2)
},
\end{eqnarray}
of the imaginary part for the first three $k\in \{0 \mbox{(solid)},2\mbox{(dashed)},4\mbox{(dotted)}\}$ partial waves of the DA, which are normalized to the full NLO result for $k=0$. The corrections are very large in the
quark-quark channel (left panel) and they grow with increasing $k$. Thereby, the pure singlet quark part reduces the $k=0$ partial wave by few percents, see dash-dotted curve, and Eq.\ (\ref{pSc_{jk}-odd}) tells us that the pure singlet quark part become strongly suppressed for higher partial waves. The gluonic contributions (right panel) are moderate, however, they grow with increasing $k$.
Note that finally the NLO corrections depend on the non-perturbative input
$\varphi^\Sigma_{\eta^{(0)},k}(\cQ_0^2)$ and $\varphi^\g_{\eta^{(0)},k}(\cQ_0^2)$,
too. From the photon-to-meson transition form factor the information
on the first Gegenbauer moment $k=2$ has been obtained
\cite{Kroll:2002nt,Kroll:2013iwa}.
%%%%%%%%%%%%%%%%%%%%%%%%%%%%
%%%%%%%%%%%%%%%%%%%%%%%%%%%%%

Finally, let us summarize. We employed an efficient and straightforward method to calculate the NLO corrections to DVMP  for the flavor singlet sector in the momentum fraction representation.
 The results were mapped into the space of conformal moments which allow in future to employ the Mellin-Barnes integral representation in phenomenology. We found that the
 NLO corrections to the pure singlet quark part  are small while the quark-gluon channel might imply moderate corrections. The main corrections are large and arise from the quark-quark channel.
\\

\noindent
{\bf Acknowledgments}

\noindent
For discussions we like to thank D.~Ivanov and J.~Wagner.   This work has been supported in part by the Croatian
Science Foundation (HrZZ) project ``Physics of Standard Model and beyond''  HrZZ 5169, the NEWFELPRO grant agreement
no.\ 54, and the H2020 CSA Twinning project No. 692194, “RBI-T-WINNING.

%\bibliographystyle{h-physrev}
%\bibliography{../../../../myTeXinput/veroefli,../../../../myTeXinput/referenc}

\end{document}